\documentclass[12pt]{article} % it's 12pt

\usepackage[hidelinks]{hyperref}
\hypersetup{allcolors=blueish,colorlinks=true}

\usepackage[compress]{scicite}
\usepackage{times}

\usepackage{amsmath}
\usepackage{amsfonts}
\usepackage{amssymb}
\usepackage{graphicx}

\usepackage{booktabs}

\newenvironment{sciabstract}{%
\begin{quote} \bf}
{\end{quote}}

\usepackage[labelfont=bf,figurename=Fig.,labelsep=period,font=footnotesize]{caption}

\title{Stronger together? The homophily trap in networks}

\author
{
Marcos Oliveira,${}^{1,2}$ 
Leonie Neuhäuser,${}^{3}$ 
and Fariba Karimi${}^{4,5}$\vspace{.2em}\\
\small{${}^{1}$Vrije Universiteit Amsterdam, Amsterdam, Netherlands}\\
\small{${}^{2}$University of Exeter, Exeter, UK}\\
\small{${}^{3}$RWTH Aachen University, Aachen, Germany}\\
\small{${}^{4}$Graz University of Technology, Graz, Austria}\\
\small{${}^{5}$Complexity Science Hub Vienna, Vienna, Austria}\vspace{.1em}\\
\footnotesize{
$^\ast$\href{mailto:m.a.oliveira@vu.nl}{m.a.oliveira@vu.nl}
$^\dagger$\href{mailto:denis.bonnay@gmail.com}{karimi@tugraz.at}}\vspace{-1em}
}

\date{}

\usepackage{changepage}

\usepackage{xcolor}
\usepackage{soul}
\definecolor{reddish}{HTML}{FBB4AE}
\definecolor{blueish}{HTML}{1F54A9}
\definecolor{magentish}{HTML}{FF00AA}
\definecolor{greenish}{HTML}{a1d99b}

\usepackage[hidelinks]{hyperref}

\newif\iflong
\longtrue

\usepackage{setspace}

\usepackage{lineno}
\begin{document} 


% \baselineskip24pt


\maketitle 

% An opening sentence that states the question/problem addressed by the research AND
% Enough background content to give context to the study AND
% A brief statement of primary results AND
% A short concluding sentence.

%\newpage
\begin{sciabstract}
Abstract.

\textnormal{%
While homophily---the tendency to interact with similar others---may nurture a sense of belonging and shared values, it can also hinder diversity and widen {inequalities}. Here, we unravel this trade-off analytically, revealing homophily traps for minority groups: scenarios where increased homophilic interaction among minorities negatively affects their structural opportunities within a network. We demonstrate that homophily traps arise when minority size falls below 25\% of a network, at which point homophily comes at the expense of lower structural visibility for the minority group. Our work reveals that social groups require a critical size to benefit from homophily without incurring structural costs, providing insights into core processes underlying the emergence of group inequality in networks.
% indicating core processes behind the emergence of group inequality in networks.
% Our work 
% The homophily trap depicts when favoring intra-group ties incurs structural costs inherited from the limitations of the group. Our work focuses on limitations regarding group size and access to opportunities, demonstrating that, in small groups, homophily comes with the cost of reduced structural opportunities.
%  Otherwise, they are in a scenario of homophily trap
% and avoid structural disadvantages.
% without compromising their visibility.
% without limiting their access to structural opportunities.
}
\end{sciabstract}
% \newpage  
% In setting up this template for *Science Advances* papers, both
% the \section* command and the \paragraph* command are used for topical
% divisions.  Which you use will of course depend on the type of paper
% you're writing.  Review Articles tend to have displayed headings, for
% which \section* is more appropriate; Research Articles, when they have
% formal topical divisions at all, tend to signal them with bold text
% that runs into the paragraph, for which \paragraph* is the right
% choice.  Either way, use the asterisk (*) modifier, as shown, to
% textsc
% \newpage
% \baselineskip24pt %% linespread
\baselineskip1.55em %% linespread

\section*{Introduction}
Homophily is ubiquitous---people tend to associate with similar others across domains of life, from education and relationships to employment~\cite{McPherson2001,McDonald2011,karimi2022minorities}. This preference for in-group ties is a trade-off: it fosters group cohesion but promotes social segregation. While homophily may nurture a sense of belonging, it can  limit access to opportunities across social group boundaries and exacerbate inequalities, particularly in networks with minority groups~\cite{dimaggio2012network,jackson_inequalitys_2021,garip2021network}. Although this trade-off is a core building block of social networks, it remains analytically unexplored.

By connecting with similar others, people skew their networks towards shared traits, reinforcing social identities and fostering a strong sense of community that provides advantages at the individual level. For example, immigrants settling in a country often leverage intra-ethnic contacts to find economic opportunities~\cite{zhou_revisiting_2004,currarini2009economic,levanon_who_2014,heizmann_migrant_2016,jackson2019human} and access emotional support networks that are less available across ethnic boundaries~\cite{McPherson2001}. This so-called in-group favoritism can further create ideal conditions for collective action. In environmental activism, for instance, activists are more likely to engage in actions when they strongly identify and interact with their groups~\cite{schulte_social_2020}. Such a preference for the familiar, however, inherently hinders diversity by segregating social groups.

This segregation can deprive individuals of accessing opportunities and participating in deci\-sion-making processes~\cite{jackson_inequalitys_2021,oliveira2022group} and can undermine the benefits of weak ties~\cite{granovetter1973strength,granovetter1995getting,au2023theoretical}, which is especially problematic in the workplace, where homophily affects recruitment directly~\cite{bushell_network_2020,fernandez_networks_2006}. For example, job referrers are more likely to recommend candidates of the same gender, race, age, and education~\cite{fernandez_networks_2006,fernandez_gendering_2005,brown_informal_2016}, leading to in-group favoritism that explains occupational segregation~\cite{buhai_social_2023}. This group preference can also contribute to a gender pay gap in healthcare, as physicians tend to refer patients to specialists of the same gender~\cite{zeltzer_gender_2020}. Beyond the workplace, group segregation can also shape how individuals perceive others depending on group affiliation, producing perception and implicit biases in networks~\cite{Lee2019,stier2024implicit}. Overall, while homophily offers benefits, it entails inherent costs.

Such a trade-off is particularly critical in networks with minorities. When minority members favor in-group ties, they inherit not only the opportunities but also the limitations of their group. For instance, at social gatherings, homophily within small social groups limits individuals' contact pool, resulting in fewer connections on average for these numerical minorities~\cite{kossinets_origins_2009,oliveira2022group}. Such inherited limitations can affect the social capital of a group~\cite{lin2002social,chetty2022social}. For example, immigrants relying on intra-ethnic contacts to find jobs might end up in low-wage positions, potentially leading to an ethnic mobility trap that hinders upward social and economic mobility~\cite{wiley_ethnic_1967,kalter_migrant_2014}. 
Despite its paradoxical effects on minority groups, the homophily trade-off remains poorly understood, lacking an analytical framework to explain its intrinsic connection to structural limitations in networks. 

In this work, we explore homophily in networks analytically to disentangle its inherent trade-off, with a focus on minorities. We investigate when homophilic ties are detrimental to minority groups, introducing the concept of the \emph{homophily trap}---scenarios where increased homophilic interaction among minorities negatively affects their structural opportunities within a network. To study these scenarios, we use a generative network model to construct networks of different group mixing and minority sizes. We show that homophily traps arise when the minority group size falls below 25\% of a network. Below this threshold, higher homophily within the minority group leads to fewer structural opportunities for the group: in-group ties come at the expense of lower structural visibility. Remarkably, this threshold holds regardless of the majority's strength of homophily, making it a universal condition under which minorities can either fall into a structural trap or tip the system to their advantage.

%By disentangling the trade-off of homophily and group size systematically, we build a foundation for understanding how homophily shapes structural opportunities in networks. 

%making it difficult for numerical minorities to both maintain a high number of connections and belong to homophilic social groups. 
\begin{figure*}[b!]
\centering
\begin{adjustwidth}{-.3in}{0in} 
\includegraphics[width=6.78in]{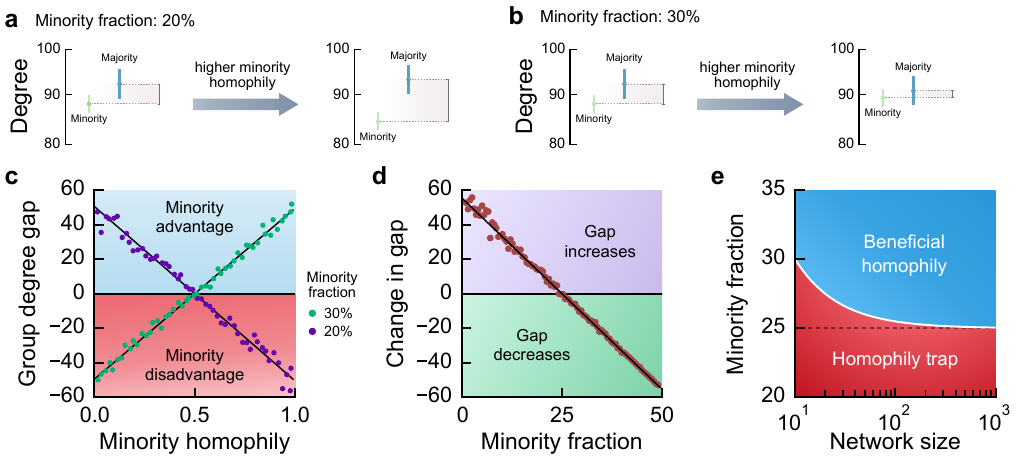}
\end{adjustwidth}
  \caption{\textbf{The homophily trap: when favoring intra-group ties incurs structural costs.} 
\textbf{(a, b)} 
  The impact of higher minority homophily on degree gap depends on group size. 
  \textbf{(c)} When the minority size is 30\% of the network, higher minority homophily benefits the group; at 20\%, however, it becomes detrimental to the group.
  \textbf{(d)} We examine this impact across different minority sizes and majority homophily values, finding that with larger minority groups, the impact of homophily shifts from increasing to decreasing the gap.
  \textbf{(e)} This shift occurs at a critical size, beyond which homophilic interaction helps the connectivity of the minority group. In large networks, scenarios of homophily trap occur when a group is smaller than 25\%, where favoring intra-group ties incurs reduced connectivity. In all plots, dots represent simulation results for networks with 1000 nodes, while curves are based on analytical formulations.}
\label{fig1}
\end{figure*}

\section*{Results}
% - P1 explain the rationale behind the model; - individual-level and relational-level attributes
To explore homophily and its implications, we focus on relational properties that affect how social groups interact in networks. Specifically, we are interested in group-level homophily, $h_{ij}$, that encodes the mixing tendency between groups, which reflects processes such as choice homophily (i.e., ties due to individual preferences), induced homophily  (i.e., ties from homogeneous opportunities such as schools or workplaces), and higher-order effects~\cite{kossinets_origins_2009,karimi2023inadequacy}. While individual-level properties, such as node degree or fitness, can affect tie formation, they have minimal impact on group-level interaction when assuming independence. Under this assumption, $h_{ij}$ determines the probability of a node from group $i$ to connect with a node from group $j$. For two groups, $0$ and $1$, we define $h_{00}$ and $h_{11}$ as the intra-group connection tendencies, with the inter-group tendencies given by $h_{01} = 1 - h_{00}$ and $h_{10} = 1 - h_{11}$.

% - P2 focus on two groups; explain the operationalization of degree ~ structural opportunities
% To investigate the effect of group mixing on structural opportunities, 
We quantify access to structural opportunities using the average degree of the nodes within each group, given that the number of connections relates to social capital~\cite{p_borgatti_network_1998,lin2002social}, and having fewer connections can result in structural disadvantages~\cite{Haas2010,Stadtfeld2019}. %such as health issues~\cite{Haas2010} and poor academic performance~\cite{Stadtfeld2019}. 
We determine the average group degrees by first finding the inter- and intra-group edge counts, which depend on relational properties and group sizes. We find that, on average, the intra-group edge counts are given by $E_{00}  =  h_{00} n_0 (n_0 - 1) /2$ and $E_{11}  =  h_{11} n_1 (n_1 - 1) /2$, where $n_i$ represents group size (see Methods for inter-group edges). With these edge counts, we determine the average group degrees $\langle k_0 \rangle$  and $\langle k_1 \rangle$, expressed as $\langle k_i \rangle = (2 E_{ii} + E_{ij} + E_{ji})/{n_i}$.

We use this expression to examine how group mixing affects average group degrees, focusing on scenarios where group $0$ is a numerical minority (i.e., $n_0 < n_1$). By varying group sizes and relational properties, we can explore the impact of increasing minority homophily on average degrees. For example, we find that when the minority makes up 20\% of the network, higher homophily decreases its average degree, whereas at 30\%, higher homophily increases it (see Fig.~\ref{fig1}a-b), indicating that the interplay of group mixing and size exhibits a transition state where homophily has double-edged sword effects. %This interplay can limit minority structural opportunities and exacerbate a degree gap among groups.

% ---number of connections relates to social capital~\cite{p_borgatti_network_1998}, and having fewer connections can result in health issues~\cite{Haas2010} and poor academic performance~\cite{Stadtfeld2019}. 

% This differences has been shown before in data. say that it implies that homophily needs to account for group size to understand its impact (nat.comm.phys. paper: The existence of these two contrasting regimes implies that the very interpretation of a minority group depends on the group mixing in the network. When studying a minority group, one has to account for inter- and intra-group dynamics to understand its position in a network.)

\subsubsection*{Group-size dependence and structural gaps}
% Given the dependencies of intra-group edges to homophily and group sizes, w
%We explore how homophily affects the structural opportunities of minority groups, showing a dependence on minority group size. Precisely, we examine the 
To investigate the role of homophily in creating structural gaps,
% we delve into our analytics and 
% Here,
we examine analytically how the impact of minority homophily varies with group size. First, we define structural gap $\Delta k = \langle k_0 \rangle  - \langle k_1 \rangle$ to represent the difference in structural opportunities between the minority and majority groups; then, we examine the impact of higher minority homophily on this gap, revealing that its effect depends on group size (Fig.~\ref{fig1}c). Specifically, increasing minority homophily can either widen or narrow the gap, depending on the relative minority size $f_0 = n_0/N$. For small minority fractions (e.g., 20\% of the network), higher minority homophily widens the degree gap, limiting the minority group structurally. 

% - P4 explain the gap change 
% 	- show the dependency on minority size (Fig. 1C)
% 	- explain the result here slowly
% 	- show the point that the gap increases (i.e., positive value) and becomes negative (i.e., negative value)
% explore how changes in minority homophily
We investigate this group-size dependence systematically to characterize the interplay between homophily and structural gaps in networks. We analyze the impact of variations in minority homophily on the degree gap across various minority fractions and majority homophily values. Our results reveal that as the minority group grows larger, the impact of a higher minority homophily shifts from widening to narrowing the gap (Fig.~\ref{fig1}d). Regardless of majority homophily values, a higher minority homophily increases the gap when the minority is small, whereas with a larger minority (e.g., 30\% of the network), the gap decreases with higher homophily. This transition represents a qualitative change in the impact of homophily on structural opportunities.

\subsubsection*{Homophily traps in networks and critical minority size}
Using the properties of such generalized network models, we analyze how the structural gap varies as a function of minority homophily through examining the derivative of the gap $\Delta k$ with respect to $h_{00}$, leading to the following expression:
\begin{equation}
\label{eq:derivative}
    \dfrac{\partial \Delta k}{\partial h_{00}} = 2 N f_0 - \dfrac{N}{2} - 1,
\end{equation}
which describes how the degree gap between minority and majority groups responds to changes in minority homophily. This derivative reveals that the effect of minority homophily on the gap depends on the size of the minority group $f_0$ and the overall network size $N$, but not directly on the homophily within the groups; that is, the derivative is independent of $h_{00}$ and $h_{11}$. 

We are particularly interested in when this derivative equals zero, as it reveals the critical minority size $f_0^*$ where the impact of minority homophily on the gap shifts the regimes from increase to decrease in the disparity. By setting Eq.~(\ref{eq:derivative}) to zero, we find that the critical minority size is given by:
\[ f_0^* = \dfrac{2 + N}{4N}. \]
This critical size depends on the network size $N$, but in larger networks, it becomes independent of $N$ and approaches a fixed value:
\[ \lim_{N \to \infty} \dfrac{2 + N}{4N} = \dfrac{1}{4} = 0.25, \]
meaning that in sufficiently large networks, the critical minority size is 25\% (Fig.~\ref{fig1}e). We validate this result through simulations, confirming the existence of $f_0^*$ in different mixing scenarios and network sizes.

When the minority group is smaller than this critical size, higher minority homophily results in wider structural gap in opportunities, leading to a case of homophily trap, which occurs due to inherent structural limits in the network. A small minority group lacks the size to create enough intra-group edges to compensate for the inter-group edges that are lost with high group homophily. Although higher homophily fosters a more close-knit minority group, these structural limits ultimately reduce broader network access for the group.

\section*{Discussion}

%While homophily fosters in-group cohesion, it implies reduction in out-group interaction, weakening the abundance of the weak ties~\cite{granovetter1973strength}, and restricting individuals' access to opportunities beyond their social groups. The trade-off between the costs and benefits of the strength of homophily and minority size remained a dilemma in the literature. 
% 
While homophily implies in-group cohesion, it reduces out-group interaction, restricting individuals' access to opportunities beyond their social groups. We provide a mathematical solution to this trade-off. We show that when a minority group is smaller than 25\% of a network, its intrinsic structural limits prevent it from both maintaining a homophilic group and attaining a high number of connections---an instance of the homophily trap. This threshold holds regardless of the majority's level of homophily, making it a universal condition under which minorities either fall into a structural trap or tip the system to their advantage.

This result demonstrates that in social networks, homophily comes with the cost of reduced structural opportunities, as measured by the group's average degree. The trap arises from the natural limits of small groups, which lack enough members to sustain homophilic ties without sacrificing connectivity. Such limits translate into a critical mass that the minority group must reach to avoid structural costs. 
  
Similar critical mass thresholds have been predicted and observed in different social contexts. In leadership, qualitative studies suggest that women must reach about 30\% representation to challenge prevailing social conventions~\cite{kanter_effects_1977,dahlerup_small_1988}. Experimental research on social coordination shows that persistent minorities can overturn established social conventions when they reach approximately 25\% of a population~\cite{Centola2018}. This particular value for critical mass has been repeatedly observed in simulation and empirical studies across disciplines, marking a tipping point where minorities can instigate social change~\cite{everall2023pareto}. 

This threshold can be read as a formal boundary condition consistent with the weak/strong ties literature: when minorities fall below a critical mass, prioritizing stronger in-group ties limits access to out-group bridges~\cite{granovetter1995getting}. It also aligns with the concept of social closure in immigrant networks, in which dense co-ethnic ties secure opportunities for insiders while restricting access for outsiders through processes of closure and segmentation~\cite{waldinger1997closure}.

Future research could explore how avoiding the homophily trap relates to fission--fusion dynamics in cultural evolution and animal societies, where individuals frequently join and leave subgroups~\cite{aureli2008fission}.%Such social structure allows for flexible responses to ecological challenges, such as resource distribution and predator avoidance~\cite{couzin2009fission}.

%Such a consistency suggests that our findings might relate to core mechanisms underlying critical mass thresholds in networks.

Our results demonstrate that homophily traps can arise from structural limits alone, even in networks without mechanisms such as preferential attachment or triadic closure. This provides a baseline for future work on how network topologies and additional mechanisms affect the homophily trade-off. By disentangling this trade-off, we build a foundation for understanding how homophily shapes opportunities and underpins group inequality in networks.

%%% some leftovers:

%Understanding the balance between homophily and diversity is crucial for addressing issues related to social inequality, perception biases, and the visibility of minority groups in various social contexts.

% \leo{The assignment of individuals to a particular group is often based on certain characteristics \cite{mcclain_group_2009}. It is important to note that the this group membership does nt imply an individual's awareness of belonging to a particular group and having a psychological attachment to it \cite{jackman_interpretation_1973}. When we work with real datasets, we often only have descriptive data about group membership available and do nt necessarily have access to data about group identification.} 
%%% 
% The homophily trap and the 25\% threshold suggest that if minorities are below 25\% and have no general knowledge of the mixing strategy of the majorities, they should avoid increasing their homophily in order to maximize the potential structural benefits that are available in networks. This naturally raises the questions whether or not minorities have learned this fact from the evolutionary stand point and how can we validate this in real-world social networks. Proper empirical validation requires experimental settings in which group strategies evolve over time and stabilize in certain parameter regions. 
%%%%

\section*{Methods}

To explore homophily and its implications, we focus on the relational properties $h_{00}$ and $h_{11}$, which represent the tendencies for intra-group connections within each group $0$ and $1$. We define the inter-group relational properties as $h_{01} = 1 - h_{00}$ and $h_{10} = 1 - h_{11}$, ensuring that the probabilities sum to one. 

To examine the degree gap between the two groups, we first calculate the average degree for each group. We express the number of nodes in each group as $n_0 = f_0 N$ and $n_1 = (1 - f_0) N$, where $f_0$ represents the relative size of the minority group. By using these definitions, we find that, on average, the inter- and intra-group edges counts can be expressed as:
\[
E_{00} = \dfrac{n_0 (n_0 - 1) h_{00}}{2}, \ 
E_{01} = \dfrac{n_0 n_1 h_{01}}{2}, \  
E_{10} = \dfrac{n_1 n_0 h_{10}}{2}, \quad \text{and}\] 
% \text{and}
\[
E_{11} = \dfrac{n_1 (n_1 - 1) h_{11}}{2}.
\]
With these edge counts, we determine the average degree for each group:
\[
    \langle k_0 \rangle = \dfrac{2 E_{00} + E_{01} + E_{10}}{n_0} \quad \text{and} \quad \langle k_1 \rangle  = \dfrac{2 E_{11} + E_{10} + E_{01}}{n_1}.
\]
Finally, we define the structural gap between groups as the difference between average degrees:
\[
\Delta k = \langle k_0 \rangle  - \langle k_1 \rangle.
\]

Next, we explore how this structural gap changes with variations in minority homophily, $h_{00}$. To accomplish this, we take the derivative of $\Delta k$ with respect to $h_{00}$:
\begin{equation}
\dfrac{\partial \Delta k}{\partial h_{00}} = \dfrac{\partial \langle k_0 \rangle}{\partial h_{00}} - \dfrac{\partial \langle k_1 \rangle}{\partial h_{00}}.\label{eq:partial}
\end{equation}
First, we differentiate each term separately:
\[
     \dfrac{\partial \langle k_0 \rangle}{\partial h_{00}} = (n_0 - 1) - \frac{n_1}{2} \quad \text{and} \quad 
     \dfrac{\partial \langle k_1 \rangle}{\partial h_{00}} = - \frac{n_0}{2}.
\]
Then, we  substitute these expressions into Eq.~(\ref{eq:partial}) and replace $n_1$ with $N - n_0$, yielding:
\[
    \dfrac{\partial \Delta k}{\partial h_{00}} = n_0 - 1 - \frac{N - n_0}{2} + \frac{n_0}{2} = \frac{4 n_0 - N - 2}{2}.
\]
Given that $n_0 = N f_0$, we can rewrite this expression as:
\[
    \dfrac{\partial \Delta k}{\partial h_{00}} = 2 N f_0 - \dfrac{N}{2} - 1.
\]
To find the minority critical size, we solve it for the case of zero $\dfrac{\partial \Delta k}{\partial h_{00}} = 0$, we find the minority critical size: 
\[
  f_0^* = \dfrac{2 + N}{4N}.
\]
This critical fraction depends on the network size $N$, but as the network size grows large, % ($N \to \infty$), 
it approaches a fixed value:
\[
 \lim_{N \to \infty} \dfrac{2 + N}{4N} = \dfrac{1}{4}.
\]
Thus, in large networks, the critical minority size is approximately 25\%.

\bibliography{references}
\bibliographystyle{naturemag}

\noindent \textbf{Acknowledgments:} 
We thank Filiz Garip and Jan Korbel for helpful feedback and discussions.
\textbf{Competing interests:}
 The author declare no competing interests.

% The authors thank James Clerk Maxwell and Albert Einstein for helpful conversations.\\
% \noindent \textbf{Funding:} This work was supported in part by the Very Generous Foundation.\\
% \noindent \textbf{Author Contributions} JAS conceived the research. JAS and JD designed the analyses. JAS and JS conducted the analyses. All authors wrote the manuscript.\\

% \paragraph*{Competing Interests}
% \noindent \textbf{Competing Interests} 
 % \noindent \textbf{Data and materials availability:} Additional data and materials are available online.

\end{document}